\documentclass[conference]{IEEEtran}
\IEEEoverridecommandlockouts
\usepackage{amsmath,amsfonts}
\usepackage{algorithmic}
\usepackage{algorithm}
\usepackage{array}
\usepackage{subcaption}
\usepackage{textcomp}
\usepackage{stfloats}
\usepackage{url}
\usepackage{verbatim}
\usepackage{graphicx}
\usepackage{cite}
\usepackage{balance}
\begin{document}

\title{An experimental evaluation of satellite constellation emulators
\thanks{This publication has emanated from research conducted with the financial support of Taighde Éireann – Research Ireland under Grant number 13/RC/2077\_P2 at CONNECT: the Research Ireland Centre for Future Networks.}
}

\author{\IEEEauthorblockN{Victor Cionca\textsuperscript{*}, Ferenc Szabo\textsuperscript{+}, Stanimir Vasilev\textsuperscript{+}, Dylan Smyth\textsuperscript{*}}
\IEEEauthorblockA{\textit{Department of Computer Science}\\
\textit{Munster Technological University}\\
Cork, Ireland\\
\textsuperscript{*}:\textit{first.last}@mtu.ie, \textsuperscript{+}:\textit{first.last}@mymtu.ie}
}

\maketitle

\begin{abstract}
  Satellite emulation software is essential for research due to the lack of access to physical testbeds. To be useful, emulators must generate observations that are well-aligned with real-world ones, and they must have acceptable resource overheads for setting up and running experiments. This study provides an in-depth evaluation of three open-source emulators: StarryNet, OpenSN, and Celestial. Running them side-by-side and comparing them with real-world measurements from the WetLinks study identifies shortcomings of current satellite emulation techniques as well as promising avenues for research and development. 
\end{abstract}


%

\section{Introduction}

Non-Terrestrial Networks (NTNs) and satellite systems have become increasingly relevant in recent years due to their potential in supporting future Wide Area Network (WAN) infrastructure. Low Earth Orbit (LEO) satellite systems are of special interest, due to their relatively lower altitude, which reduces path loss and latency, and thereby lowers per-link cost and infrastructure complexity compared to higher orbits~\cite{wang2025non}. Complexities surrounding the use of satellites and LEO constellations for network connectivity has led to new research questions around topology control, mobility, routing, resource allocation, and resilience. The need for researchers to evaluate work aimed at addressing the challenges of LEO and NTNs usage have led to the creation of satellite simulators and emulators.

Satellite system simulators and emulators aim to provide a low-cost and accessible means of experimentation and evaluation of work in the NTN domain. Simulators provide scalability, while emulators provide a realistic software stack on each emulated satellite and Ground Station (GS) nodes. A common feature of both tool types is simulating the orbit of the satellite nodes and maintaining accurate link characteristics for each node based on their relative positions around the globe. The goal is to generate behaviour that is consistent with real-world observations, which is known as \textit{fidelity}. Existing simulation and emulation tools use a variety of technologies to implement deployment, networking, software stacks, and experiment data recovery. The methods employed introduce trade-offs in the fidelity and performance of the tool.

Increasingly networks researchers are leaning towards emulation, where the added realism can capture systems-level behaviour that simulation abstracts away. This paper evaluates and compares the performance and feature implementation of three satellite system emulators: StarryNet ~\cite{lai2023starrynet}, OpenSN ~\cite{lu2024opensn}, and Celestial~\cite{pfandzelter2022celestial}. These tools were chosen as 1) they capture all the characteristics of a satellite network, with both Ground-Satellite (GSL) and Inter-Satellite links, and 2) their source code is readily available with support still provided by their authors. The design and implementation of each tool is first detailed (\S\ref{sec:solutions}), followed by an evaluation of each tool's fidelity and performance at scale (\S\ref{sec:evaluation}). Fidelity is measured by comparing the tool's output with real-world measurements from the WetLinks study~\cite{wetlinks}. Performance at scale is evaluated through CPU usage and delays incurred when bringing up experiments and updating the network characteristics of the simulated/emulated NTNs. 

The study concluded that:
\begin{itemize}
     \item none of the existing satellite constellation emulators can match real-world measurements because important factors such as client scheduling or atmospheric conditions are not considered
     \item constellation update delays are not bounded, however, while the experiment bring-up time is acceptable, constellation update delays are large and often cause disruption during emulation.
\end{itemize}
The observations collected from the study (\S\ref{sec:observations}) provide guidance for future implementations in terms of useful and required features.





\section{Satellite networking simulation and emulation}
\label{sec:solutions}

Network experimentation has the requirements of scalability and realism: researchers must be able to build large networks that exhibit similar at-scale behaviour to real networks;  they also need to capture real system effects such as OS overheads, physical issues, etc. Without access to actual hardware infrastructure the two tools available are simulators and emulators, which balance the two requirements: simulators can achieve large scale at the expense of realism, while emulators have higher realism at the expense of scale. Emulation is more attractive because it allows the evaluation of real-systems directly, without having to translate system logic into the reduced functionality environment of a simulator. Network emulation is typically implemented through virtualization. The overhead of virtualization has been reduced by technologies such as containerisation and microVMs~\cite{agache2020firecracker}, and horizontal scaling over multiple servers has virtually eliminated the problem.

There are limitations to using emulators: primarily because they rely on virtualization technology they emulate traditional computers with Ethernet-based network interfaces. Emulation of custom hardware or custom Layer 2 networking requires additional software support, and that is where simulators are useful. In satellite networks the ground-space radio channel suffer the combined effect of: atmospheric conditions, local obstructions, antenna configuration and beam-forming~\cite{neinavaie2022unveiling}, as well as multi-access scheduling~\cite{tanveer2023making}. It is possible to represent some of these as an average packet loss or delay that can be reproduced in emulation, however simulators can achieve increased accuracy.

A brief review of satellite simulators and emulators is provided before detailing the emulators evaluated, StarryNet~\cite{lai2023starrynet}, OpenSN~\cite{lu2024opensn}, and Celestial~\cite{pfandzelter2022celestial}.

\subsection{Simulators}

Initial focus of satellite simulation was on capturing the mobility patterns and developing accurate models of the ground-space communication channel that include the impact of weather and other atmospheric conditions~\cite{niehoefer2013cni}. SNS3~\cite{puttonen2015satellite} was one of the first simulators to capture packet-based communication for geostationary multimedia satellites. With the rise to prominence of LEO constellations simulators such as Hypatia~\cite{kassing2020exploring} and StarPerf~\cite{lai2020starperf} were developed, focusing on the scale requirements of the large constellations, on the complex communication within the constellation, and on the dynamic ground-space connections. Recently the hosting of services (e.g. CDNs) inside LEO satellites has been considered, and some simulators such as Stardust~\cite{pusztai2025stardust} have focused on evaluating the placement and access to services hosted in LEO networks. A recent in-depth review of satellite simulation is provided by Manzanares-Lopez et al. in \cite{manzanares2025comprehensive}.

Hypatia~\cite{kassing2020exploring} is a LEO satellite network simulation framework that pre-computes satellite positions and network states over time, enabling efficient modelling of large-scale constellations. It integrates with the ns-3 simulator to perform packet-level network simulations and supports interactive 3D visualizations through Cesium, helping to analyse and understand satellite network dynamics, connectivity, and performance. It supports ISLs in grid model, supports more accurate channel models (through ns-3), however does not implement the scheduling of multiple ground stations when talking to the same satellite.

StarPerf~\cite{lai2020starperf} takes a different approach to Hypatia, aiming to provide average estimates for latency, throughput and coverage of a mega-constellation without exchanging network packets. Given the description of a constellation that includes routing policy (shortest path or source-routing) it builds a graph representing areas of interest and calculates latency and throughput as a function of link and path length on the graph. It can also run further graph-theoretic analyses such as network resilience. Without packet-level simulation StarPerf can achieve larger scale than Hypatia, however is unable to capture effects of network load.

A recent solution is Stardust~\cite{pusztai2025stardust} that considers a "3D continuum" consisting of ground, satellite, and intermediate networks (e.g. drones, balloons) and allows users to build simulations of computational tasks run over the networks. Stardust implements a link protocol that maintains links between nodes, and a network protocol that performs routing. Computation is user-defined and the evaluation shows that a network of more than 20,000 nodes can be simulated on a 32-core machine.

\subsection{Emulators}

In network emulation, each node is ideally a fully working system with an operating system and suite of user-space tools. This is achieved in most cases using virtualization technologies with low overhead such as containers and microVMs. The nodes are then interconnected using Linux bridges and virtual adaptors~\cite{splitnn}. There are variations to this template meant to reduce resources requirements or reduce the experiment bring-up time.

Satellite emulators are different from traditional networks because of the deterministic mobility of the nodes. As such, emulators will use satellite modelling software such as STK~\footnote{https://www.ansys.com/products/missions/ansys-stk} or GMAT~\footnote{https://software.nasa.gov/software/GSC-17177-1} to periodically compute satellite positions and the delay and packet loss of links. The link characteristics can be enforced onto the virtual links using the Linux \texttt{tc} and \texttt{netem} tools~\cite{linuxtc} or similar.


One of the earliest satellite emulators is LeoEM~\cite{cao2023satcp} which is open-source and allows low-resource emulation of satellite constellations. It uses Mininet~\cite{huang2014teaching} to represent network nodes and to manage the links between them. However it is imperative to observe that LeoEM was not designed as a generic satellite emulator but with the intention of evaluating a TCP algorithm. For that reason, LeoEM does not emulate an entire constellation but only a path through the constellation. There is no routing and changes on the path are passed to the user applications as downtime.

A similar approach is $x$eoverse~\cite{kassem2024x}, that also uses Mininet to implement the constellation, and furthermore simulates the impact of weather conditions on the links. $x$eoverse precomputes constellation conditions, as well as best routes through the constellation, however it is not clear how the links are configured, whether routing can be left up to the user or $x$eoverse simplifies the connectivity based on the precomputed routes. It is notable that $x$eoverse addresses the same issues discussed in this paper, namely the need for emulators that are realistic and responsive at any scale. 

Rhone~\cite{wang2025emulating} is a recent solution that focuses on hardware realism with models for harvested energy and computational performance as impacted by temperature variation. The emulation techniques are same as StarryNet~\cite{lai2023starrynet}, with the listed models used to control the computational resources allocated to each container; furthermore, each satellite can now be represented by a set of containers capturing different modules in the satellite (e.g. CPU  + GPU). However, at the time of writing Rhone was not open-source.

SplitNN~\cite{splitnn} is a recent solution for enabling emulation of large scale networks, with an application on LEO constellations. The authors show that the traditional method for instantiation of virtual links, where virtual Ethernet adapters are associated with a single network namespace, suffers from significant overhead in the Linux kernel that results in super-linear scaling. They propose instead the partitioning of virtual links between multiple OS kernels through virtual machines, and within the VMs, further partitioning in multiple network namespaces. This results in significant reductions for network bringup time.

\begin{figure*}
    \centering
    \begin{subfigure}[t]{0.3\linewidth}
        \includegraphics[width=\textwidth]{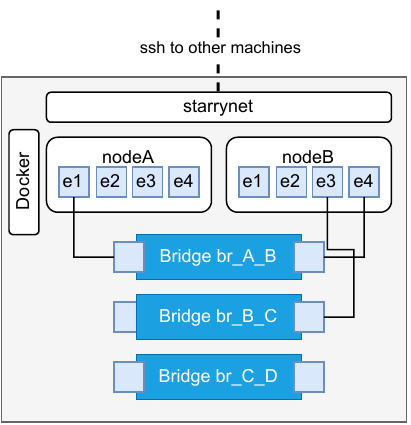}        
        \caption{StarryNet}
        \label{fig:starrynet_arch}
    \end{subfigure}
    \begin{subfigure}[t]{0.3\linewidth}
        \includegraphics[width=\textwidth]{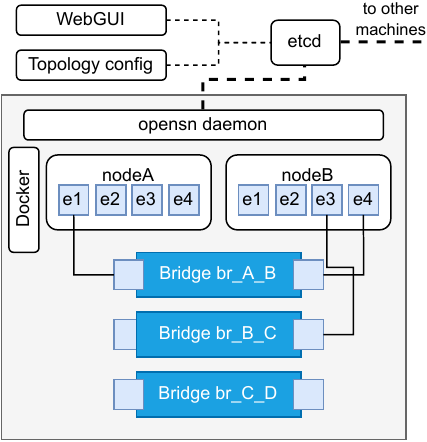}        
        \caption{OpenSN}
        \label{fig:opensn_arch}
    \end{subfigure}
    \begin{subfigure}[t]{0.3\linewidth}
        \includegraphics[width=\textwidth]{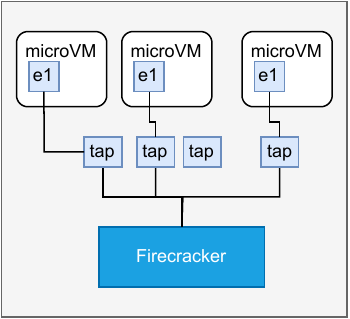}
        \caption{Celestial}
        \label{fig:celestial_arch}
    \end{subfigure}
    \caption{Architecture for the evaluated satellite emulators}
    \label{fig:architecture}
\end{figure*}

The following subsections provide in depth details about the three open-source emulators evaluated, StarryNet~\cite{lai2023starrynet}, OpenSN~\cite{lu2024opensn}, and Celestial~\cite{pfandzelter2022celestial}. A high level architecture for each of them is shown in Fig.~\ref{fig:architecture}, highlighting technologies used for emulation of nodes and networks, as well as any additional components; these are discussed below.

\subsection{StarryNet}
\label{sec:starrynet}

StarryNet, was developed by Lai \textit{et al.}~\cite{lai2023starrynet} to emulate "mega" constellations with spatial and temporal consistency and provide system-level and networking stack realism, allowing users to run real instead of simulated applications and system components. The authors evaluate the fidelity in terms of throughput and latency when compared to real measurements~\footnote{https://forschung.fh-kaernten.at/roadmap-5g/en/2021/07/24/analysis-of-a-starlink-based-internet-connection/} and existing simulators (Hypatia~\cite{kassing2020exploring} and StarPerf~\cite{lai2020starperf}). They report the resource utilisation and overheads for setting up and managing the emulation.

StarryNet is open source and currently has a mainstream as well as a development version, with significant differences between the two. The mainstream version uses Docker containers organised in a swarm for emulating the constellation and ground station nodes. Connectivity is set up using virtual Ethernet adapters and Linux bridges, using the Docker interface, and StarryNet uses the \texttt{tc netem} framework~\cite{linuxtc} to manage link delay and packet loss. All management of containers and networks is done through Command Line Interface (CLI), and this, as will be shown, introduces significant delays. The constellation configuration is defined manually with parameters such as altitude, inclination, number of orbital planes and number of satellites. Internally StarryNet uses the "Simplified General Perturbations" model (SGP4) and Skyfield Python API~\footnote{https://rhodesmill.org/skyfield/} to compute satellite positions. These, as well as the network links and their characteristics are calculated \textit{before} the emulation starts \textit{for the each step of the emulation}. This is a lengthy step because the computations are complex; however it significantly reduces delays at emulation time when StarryNet simply reads configurations from the file corresponding to the current emulation step. Such emulation is termed "trace-driven". At emulation time StarryNet will
\begin{itemize}
    \item periodically (user-defined) recalculate the link delays
    \item whenever the constellation topology changes update the links, adding and removing links as required.
\end{itemize}
Mainline StarryNet can be spread over several machines through the Docker swarm. A single binary is run on a management machine and this connects over SSH to emulation machines, executing Docker commands. The software run on the emulated satellites and ground stations consists of a hard-coded container that automatically runs the \texttt{bird} routing engine~\footnote{https://bird.network.cz/}, and has most of the essential networking tools such as \texttt{iperf3}, \texttt{ping}, or \texttt{traceroute}.

The development branch of StarryNet addresses the performance shortcomings of the mainstream version as follows:
\begin{itemize}
    \item uses the Linux Netlink API for creating and configuring network interfaces, instead of using CLI
    \item abandons Docker in favour of a custom container runtime based on Linux namespaces.
\end{itemize}
With the two changes, bringing up the constellation as well as updating links is significantly faster than the original version. However, using custom containers instead of Docker comes at the cost of portability: the new containers do not have a root filesystem, relying instead on that of the host. The host must then have all the tools required for the experiment. It is not known if and how the improved system deals with constellations of heterogeneous nodes. Of course users can generate their own root filesystems and mount that in the containers, however at that point one can argue that a dedicated solution such as Docker, Kubernetes, or podman might be more suitable. Finally, it must be said that the development branch of StarryNet is not yet considered stable by its developers.

\subsection{OpenSN}
\label{sec:opensn}

OpenSN, by Lu \textit{et al.}~\cite{lu2024opensn} uses the same technologies as StarryNet, with Docker for node emulation and Netlink for virtual links. There are some significant differences, some of them providing performance gains, others losses. First, OpenSN uses Docker and Netlink APIs instead of CLI for constellation emulation. This leads to significantly faster constellation bring-up time and link updates. On the other hand, OpenSN chooses to calculate satellite positions and other characteristics at runtime, where StarryNet precomputes them. This introduces significant delays at runtime that affect the emulation, desynchronising the satellite constellation state from the real-time measurements. 

Another major difference from StarryNet is the use of a more elaborate software architecture. Instead of using a monolithic architecture where the satellite emulation and constellation state updates are closely coupled, as does StarryNet, OpenSN disconnects the two components using a key-value store (etcd~\footnote{https://etcd.io/}) and event-based communication between components. The result, shown in Fig.~\ref{fig:opensn_arch} is a disaggregated architecture consisting of: the constellation state manager, that defines the constellation topology, number of orbital planes, etc., and periodically updates the state, pushing the values into the key-value store; the second component is the emulation daemon that receives constellation updates from the key-value store and interfaces with Docker and Netlink to apply them to the containers and virtual links. The disaggregation has the advantage that the constellation state manager can now be run on a dedicated machine where vertical scaling can be used to match the resource requirements, and there is no interference from the emulated containers. The opposite happens for the emulation daemon, which must now run on each emulation machine and consumes resources that should be used by the emulated containers.

In terms of constellation configuration, OpenSN uses the standardised TLE (Two-Line Element) notation, which should allow it to directly import configuration of real satellite constellations. Code is provided to generate TLE data for custom constellations. A 3D visualisation tool of the constellation overlaid on the globe is a very useful feature allowing users to develop custom scenarios. The containers used for satellites and ground-stations can be configured by the user, with the default ones running the \texttt{frr} routing engine~\footnote{https://frrouting.org/} and most essential networking tools.

Ultimately OpenSN provides significant performance improvements over the mainline StarryNet simply by using APIs instead of CLI. The distributed architecture should allow more flexibility in allocating resources. However, the use of runtime updates of the constellation state introduces large delays at emulation times that can affect the fidelity of the measurements.

\subsection{Celestial}
\label{sec:celestial}

Celestial, by Pfandzelter \textit{et al.}~\cite{pfandzelter2022celestial} takes a very different approach to StarryNet and OpenSN. It uses microVMs instead of containers, based on the Firecracker hypervisor~\cite{agache2020firecracker}. Each VM requires a root filesystem as well as a kernel, and Celestial provides a solution for building the root filesystem for customisation. The network emulation is also different. Nodes (satellites and ground stations) have a single network interface that is used to communicate with the hypervisor. Each microVM has its own subnet consisting of the microVM interface and the endpoint on the hypervisor, the latter acting as a gateway. The hypervisor connects to all microVMs and works as a router, therefore each microVM can potentially talk to all other ones. Celestial enforces blocking and unblocking of links either through \texttt{eBPF} filters or \texttt{iptables} rules.

Other features support the user in creating custom scenarios. The user can define:
\begin{itemize}
    \item constellations with multiple shells
    \item constellations where the arc of the ascending nodes is different from $2*\pi$, and the longitude of the first orbital plane can be offset; this is useful for simulations that use only a subset of a real constellation due to reduced availability of computational resources
    \item bounding boxes over the globe such that VMs for satellites that are outside the box are suspended and don't consume resources. An example is testing trans-Atlantic communication only in the Northern hemisphere. This feature facilitates the emulation of large scenarios even on reduced resources.
\end{itemize}
Constellation configuration is done manually, similar to StarryNet, by specifying the parameters such as number of orbital planes, etc. The constellation state is precomputed, with the emulation being trace-driven, same as StarryNet. A visualisation tool can take the constellation configuration and animate it, which, similarly to OpenSN's visualisation tool, helps the user set up custom scenarios. The approach is however different and more effective than OpenSN since it occurs separately from the emulation.

On the surface, the use of microVMs instead of containers would result in increased resource consumption. However, the bounding box mechanism means that in most cases less than half of the satellites around the world would actually consume resources, and experimental results support that. Only scenarios interested in constellation-wide routing would not benefit from the bounding box. Furthermore, the ability to run a dedicated kernel can enable more complex emulations, with custom network stacks, or hardware models.

Celestial's solution for the virtual network provides further reduction of resource consumption and delays, however it is achieved at the expense of fidelity. The constellation network, which would normally have a grid topology is effectively reduced to a star topology, as explained above, with each node microVM connected through a single virtual adapter to the hypervisor. From the point of view of overheads this means that there are four times fewer links to manage. The downside is that Inter-Satellite Links and multi-hop packet transmission through the constellation cannot be accurately emulated. Normally satellites would run routing engines that broadcast or multicast state to their neighbours, and that propagates through the constellation allowing the nodes to discover paths. In Celestial, satellites have only the hypervisor for neighbour. Satellites would find that they can reach any other satellite in two hops.

Table~\ref{tab:comparison} compares the open-source emulators in terms of their emulation technologies and features that assist scenario configuration.
\begin{table*}[]
    \centering
    \begin{tabular}{r|c|c|c|c}
              & StarryNet  & StarryNet-dev & OpenSN     & Celestial \\
        \hline
        Node emulation & Docker CLI & custom        & Docker API & Firecracker \\ \hline
        Nwk topology & grid & grid         & grid       & star \\ \hline
        Link up/down  & CLI  & Netlink API  & Netlink API & none \\ \hline
        Link properties & CLI & eBPF & tc netem & eBPF \\ \hline
        Constellation state & precomp & precomp & runtime & precomp \\ \hline
        Custom constellations & N & N & N & Y \\ \hline
        Visualisation & N & N & runtime & offline \\ \hline
        Bounding box & N & N & N & Y \\
        \hline \hline
        \end{tabular}
    \caption{Comparing the satellite emulators}
    \label{tab:comparison}
\end{table*}

\section{Evaluation}
\label{sec:evaluation}

Satellite emulators have two components, node emulation and constellation state updates, as explained in Sec.~\ref{sec:solutions}, and they function in two different time frames. The emulated nodes and the links that interconnect them represent the real-time. The node software, even though virtualized, runs on the underlying hardware CPUs; network measurements such as throughput or latency are subject to the timings of the hardware CPU, within the limits of the emulation machine~\footnote{Unless the emulation machine is running a real-time kernel timing requirements cannot be enforced. As system load increases it is possible that virtualized software experiences higher delays.}. On the other hand, the management of the constellation network, i.e. setting up and tearing down of links and updating their characteristics is simulated: it can happen in a different time period (e.g. to align with a measurement study), at a different time resolution (e.g. 1 min per simulation step, to speed up), and updates are calculated either periodically, or subject to events, or both, as is the case for StarryNet~\cite{lai2023starrynet}.

The above shows that existing solutions contain a combination of real-time and simulation-time components. To achieve a correct and meaningful evaluation of a real-world scenario the emulation solution must align the two timelines, otherwise there will be inconsistencies. Consider a network measurement of Round-Trip Time (RTT) that in the real world should be interrupted by a link change, which would result in increased values. If the simulation time is delayed the link change would occur after the measurement, yielding inaccurate values.

The three satellite emulators considered are evaluated to measure their fidelity as well as their resource requirements and runtime overheads. The measurements below were conducted for a LEO constellation inspired by Starlink, with only one shell using 53$^\circ$ inclination, 72 orbital planes and 22 satellites per plane. Where possible subsets of the orbital planes were used to reduce resource requirement.


\subsection{Fidelity}
\label{sec:fidelity}

The fidelity of the emulators is evaluated by comparing network measurements collected on the emulators with real-world network measurements published by the WetLinks~\cite{wetlinks} study.
WetLinks~\cite{wetlinks} was a measurement study of Starlink network performance conducted over 6 months from October 2023. The study collected throughput, latency, and route information from two sites in Osnabr\"{u}ck, Germany (approx. 80000 data points), and Enschede, The Netherlands (approx. 60000 data points), over a Starlink network (using the 53$^\circ$ orbit), to a nearby ground station, and to a common measurement server. The network measurements are paired with weather measurements collected from the two sites, with interesting correlations made.

The emulators were set up to replicate the WetLinks study, focusing only on the Osnabr\"{u}ck site, since there is no systematic difference between the two sites:
\begin{itemize}
    \item Where possible the experiments replicate the date "2023-09-15" of the study, which has the highest solar radiation and lowest humidity. The selection is meant to minimise the impact of atmospheric conditions on the communication as the emulators don't consider them. 
    \item The 53$^\circ$ orbital plane is used. To reduce the resource requirements only a subset of the orbits are used (10 out of 72) where possible, maintaining the angular distance between orbits and ensuring that they cover the two sites throughout the experiment, while still introducing handovers. For OpenSN we use the subset of orbits starting at longitude 225$^\circ$; for StarryNet we start at 180$^\circ$. The difference is due to the different constellation configuration techniques, the more constrained TLE for OpenSN versus manual configuration for StarryNet.
    \item 22 satellites are deployed on each orbit.
    \item Two ground stations are emulated: one corresponding to the user-terminal site in the Osnabr\"{u}ck campus, at lat. 52.28375864272186, long. 8.031676892719231; the other corresponding to the closest ground station in the vicinity of the town of Aerzen, owned by the Onlime telecoms provider, at lat. 52.06076175017756, long. 9.329243738284163.
\end{itemize}

\begin{figure}
\centering
\begin{subfigure}[t]{0.45\textwidth}
    \includegraphics[width=\textwidth]{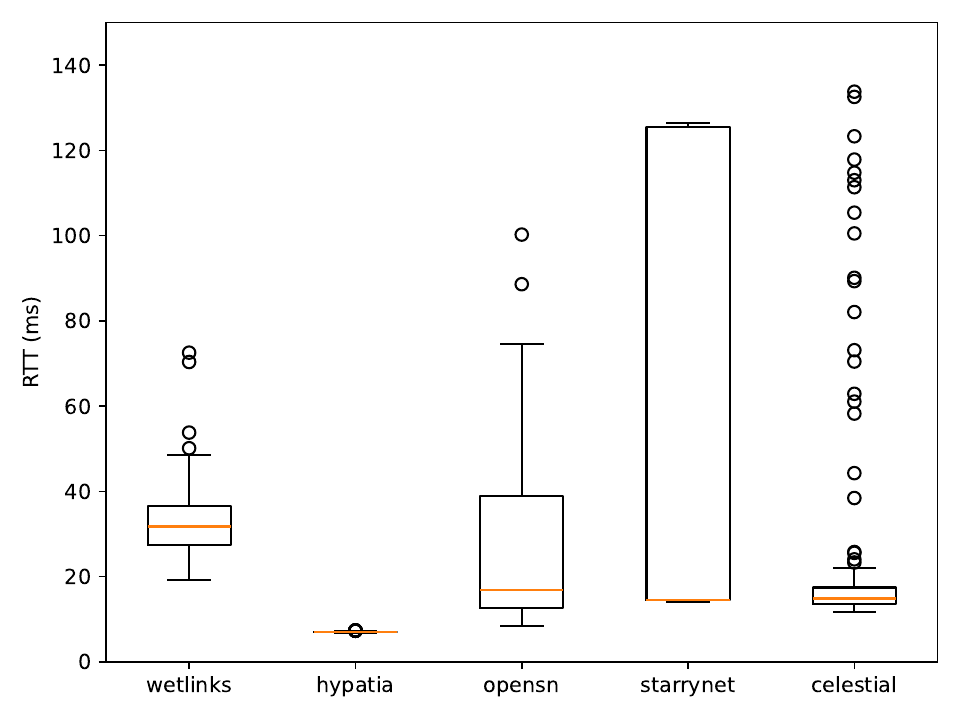}
    \caption{Bent-pipe Round-Trip Time, compared to WetLinks data}
    \label{fig:rtt}
\end{subfigure}\\
\begin{subfigure}[t]{0.45\textwidth}
    \includegraphics[width=\textwidth]{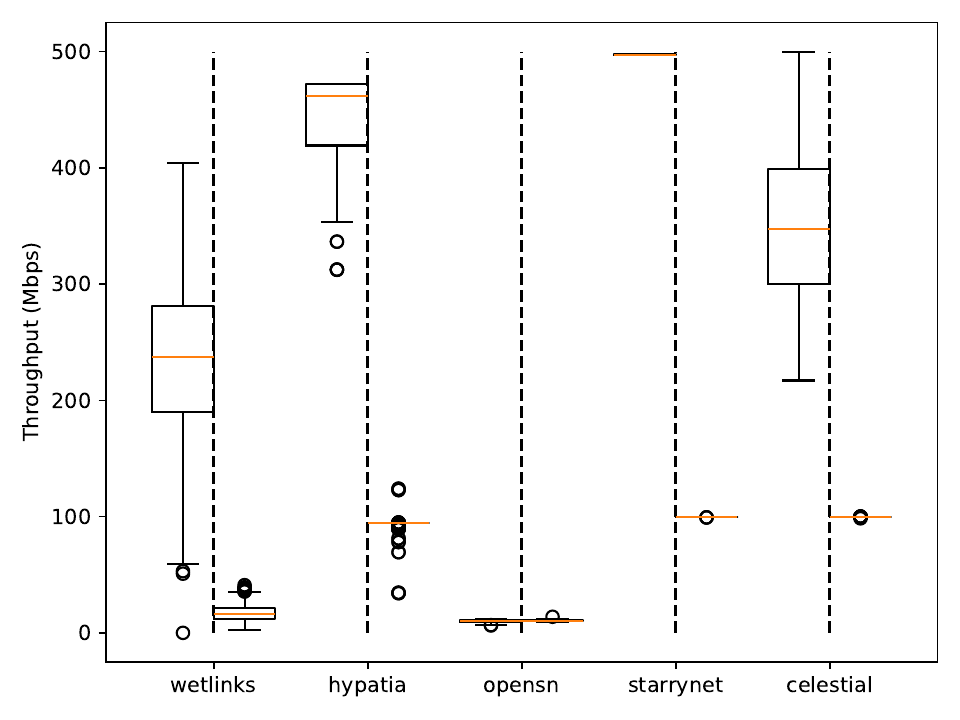}
    \caption{Uplink (left of dashed line) and downlink (right of line) throughput, compared to WetLinks data}
    \label{fig:throughput}
\end{subfigure}%
\caption{}
\end{figure}

Measurements of throughput and latency are conducted between the two ground stations with \texttt{iperf3} and \texttt{ping}, in the same manner as the WetLinks study: every three minutes two simultaneous \texttt{iperf3} sessions, one in uplink (limited at 100Mbps), the other downlink (limited at 500Mbps); this is followed by a \texttt{ping} session with 250 ICMP requests at 0.1s interval. The collected measurements are compared with the throughput and \textit{bent-pipe} latency reported by the WetLinks study. The bent-pipe path is client-satellite-ground station. The latency measured as Round-Trip Time (RTT) is shown in Fig.~\ref{fig:rtt}; throughput is shown in Fig.~\ref{fig:throughput} with downlink on the left side of the vertical dashed lines, and uplink on the right.

The first observation is that none of the results obtained with the emulators is even close to the real-world measurements, there is no need for statistical comparison. The likely cause is the simplistic ground-space connectivity model that the emulators use. The impact of atmospheric conditions is emulated through a fixed average packet loss, and transmission delay is based on free-space propagation. Satellite selection which has been shown to follow complex policies in real constellations~\cite{tanveer2023making} is done based on closest distance in the emulators. Finally, none of the emulators implement the multi-access scheduling of multiple ground clients simultaneously talking to the same satellites. Even though previous works assume that this should not be needed due to careful frequency allocation~\cite{kassing2020exploring} it has been shown to exist in Starlink satellites, with an important impact on RTT~\cite{tanveer2023making}.

In the real world, RTT measurements contain variation caused by: i) clients switching satellites as they lose connection; ii)  MAC layer scheduling of multiple clients to the same satellite; iii) weather conditions. The emulators capture the handovers between satellites, and introduce constant packet loss, however they do not capture scheduling or weather effects. The same conditions can be observed in the real-world throughput measurements. As per the WetLinks study, the uplink test was limited at 100Mbps, with the downlink at 500Mbps. StarryNet achieves those values with very little variation. Celestial produces some variation in the downlink session, bringing it closer to the WetLinks measurements. OpenSN achieves very close values in the uplink to WetLinks, however the downlink values are very low. We believe this to be an issue with the configuration of link rate. By default, OpenSN configures all links with a 1.5\% packet loss, while StarryNet uses 1\% packet loss.

The evaluation also considers the ISL support by measuring RTT on a multihop path between a EU and a US groundstations, specifically, the OnLime groundstation used above, and the Triunfo Pass Earth Satellite Station in California at latitude 34.0810947 and longitude -118.8991708. With OpenSN the path takes 9 hops and approx. 100ms. The path built by StarryNet was very unstable with 10 hops and between 200-300ms RTT. In Celestial the path takes 80-90ms RTT however the path is only 2 hops. This is because Celestial employs a star topology as discussed above, therefore it does not properly emulate ISLs. The Hypatia results are similar to Celestial with values between 86-99ms. Considering that the distance around the world between the two locations is $\approx9200$km, free-space propagation time would be about 30ms, so 60ms RTT. The delay at each intermediate hop is usually sub-ms~\cite{papagiannaki2003measurement}, therefore neglijible compared to the propagation time. Propagation through the ISLs will not be on the shortest path due to the grid nature of the constellation topology; further delays will be introduced by handovers. Therefore the values of 100ms are satisfactory.

\subsection{Resource requirements}
\label{sec:requirements}

\begin{figure}
\centering
\begin{subfigure}[t]{0.45\textwidth}
    \includegraphics[width=\textwidth]{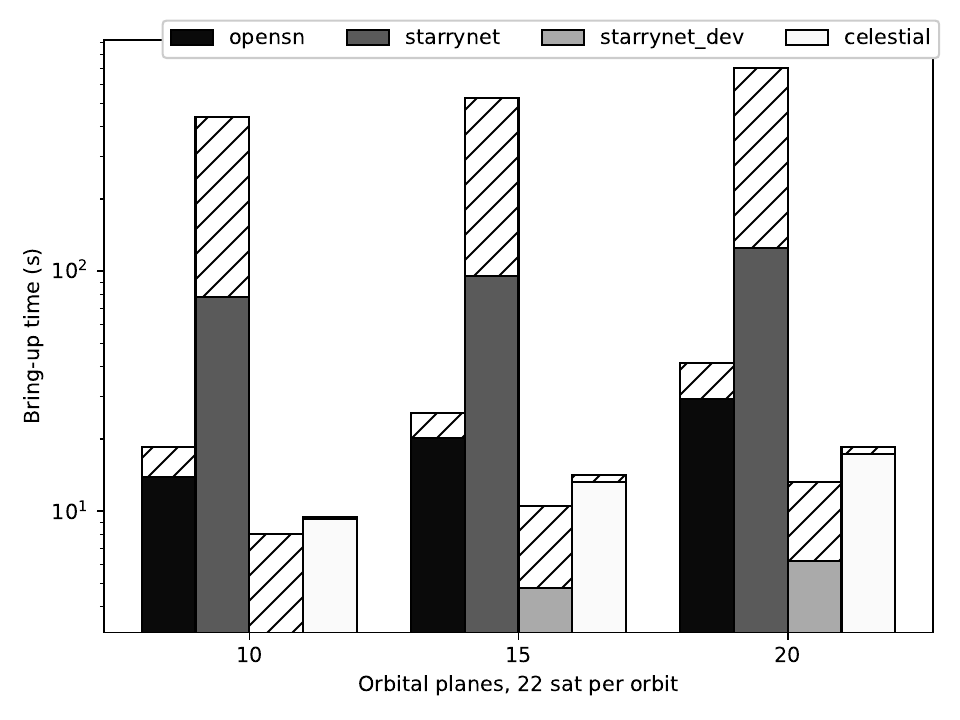}
    \caption{Bring-up time. Solid bars (bottom) -- nodes; hashed bars (top) -- networking.}
    \label{fig:bringup}
\end{subfigure}\\
\begin{subfigure}[t]{0.45\textwidth}
    \includegraphics[width=\textwidth]{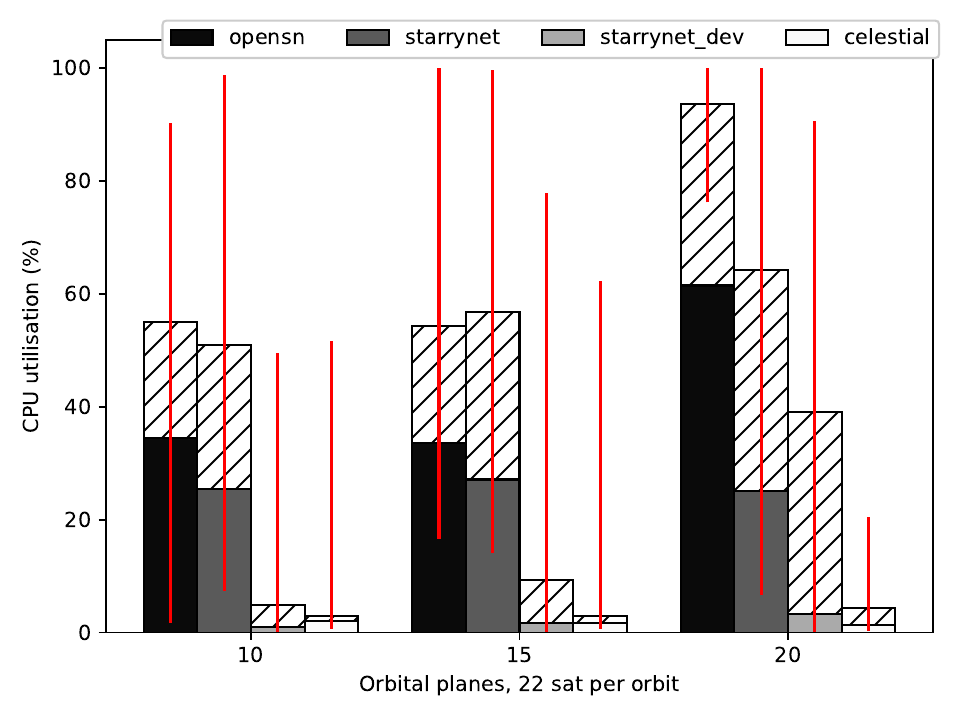}
    \caption{CPU utilisation. Solid bars (bottom) -- user-space; hashed bars (top) -- kernel.}
    \label{fig:cpu_utilisation}
\end{subfigure}
\caption{}
\end{figure}

When emulating large satellite constellations it is important to understand the resource requirements, and how they scale. On one hand, emulation requires CPU time and memory to run the software in each emulated node, and additional resources for the virtual networks. Another important resource that is affected by scale is the time required to bring up the constellation, creating and starting all the node instances, and enabling and configuring all the links. The publications corresponding to the three satellite emulators each report the values listed above, with different hardware and scenario configurations therefore not allowing conmparison. The following provides a comparison between OpenSN, StarryNet, and Celestial obtained while running them on the same configuration: the constellation is deployed on a 16-core machine with 32GB of RAM, with the management interface run on a separate machine with 4 cores. The comparison also includes the development version of StarryNet. The emulators are running, per emulated node, a routing engine (\texttt{frr} or \texttt{bird}), and periodically \texttt{iperf3} and \texttt{ping} sessions, as explained above. Celestial does not run a routing engine for the reasons stated above (\S\ref{sec:celestial}). We report the bring-up time in Fig.~\ref{fig:bringup}, consisting of initialising nodes and networking, as well as CPU utilisation in Fig.~\ref{fig:cpu_utilisation}, for constellations consisting of 10, 15, and 20 orbital planes, each with 22 satellites. 

The bring-up time shown in Fig.~\ref{fig:bringup} distinguishes between nodes and networking as the two are initialised in sequence by the emulators. The mainline StarryNet experiences almost 10x longer bring-up times than the other solutions. Even though it uses Docker containers as OpenSN, it interacts with Docker through CLI which introduces the additional overheads. The solution is not practical and it was addressed in OpenSN by using the Docker APIs for container interaction, and the Linux Netlink APIs for creating and configuring network interfaces. Bringing up the Firecracker microVMs in Celestial is faster than the Docker containers in OpenSN however: Celestial implements a bounding box on the emulation, with microVMs corresponding to satellites outside the box being suspended. Celestial therefore has microVMs in several states (created - started - suspended), and the initial bring-up time only represents the creation of the VM, with the starting of a VM taking 2-300ms. Therefore the comparison between OpenSN and Celestial bring-up is not entirely accurate, because there is a difference in number of node instances concerned. Finally, the development version of StarryNet experiences the fastest node bring-up time, mainly because of the custom container runtime and because it does not use a dedicated filesystem for each node. We believe that using a dedicated root filesystem (which would facilitate reproducibility, management of dependencies, etc.) would raise the bring-up time closer to that of OpenSN.

With regards to link bring-up time (hashed bars in Fig.~\ref{fig:bringup}), Celestial has very little overhead because it creates a star topology therefore the number of links is much lower than the other solutions. StarryNet-dev and OpenSN both use the Netlink API to configure links, therefore it is interesting that StarryNet-dev scales better than OpenSN. The slow-down experienced by OpenSN could be caused by contention for the CPU: as can be seen in Fig.~\ref{fig:cpu_utilisation}, at 20 orbital planes OpenSN reaches almost 100~\% utilisation.

The CPU utilisation shown in Fig.~\ref{fig:cpu_utilisation} breaks down utilisation into user-space (bottom bars, solid) and kernel-space (top bars, white, hashed). The utilisation is obtained from the Linux "top" tool and the kernel-space value includes kernel processes, waiting, hardware and software interrupts. Celestial experiences the lowest CPU utilisation, despite using VMs, which generally have higher resource consumption than containers. However, it is important to note that Celestial does not run routing protocols therefore effectively each VM is idle most of the time (except when performing measurements). Next is the development version of StarryNet, which uses the custom runtime based on network namespaces. Without the overhead of container management daemons such as Docker, StarryNet-dev achieves much lower CPU utilisation than OpenSN and the mainline StarryNet. Nevertheless the kernel overhead for managing the containers increases until at 20 orbits (440 satellites) it is comparable to that of OpenSN and mainline StarryNet. Finally, OpenSN and mainline StarryNet, working with Docker containers present the highest utilisation, with OpenSN scaling the worst, reaching almost 100\% at 20 orbital planes. It is interesting to note that the user-space utilisation of StarryNet is stable. This is because it represents the overhead of the container manager (one for all the nodes), and little overhead from the routing engines running on each node. On the other hand, OpenSN shows much higher user-space CPU utilisation, which is caused by running the OpenSN daemon, whose resource requirements increase with the constellation size. Remember that StarryNet and Celestial have the emulation daemon completely separated, whereas OpenSN runs the daemon on each machine (\S\ref{sec:opensn}).

\subsection{Constellation update delays}
\label{sec:updates}

An important part (if not the most important part) of a satellite constellation emulator is representing the dynamic characteristics of the constellation: positions of the satellites, connectivity between satellites, link characteristics such as delay and loss, etc. When running any sort of measurements on the emulator, such as network performance, these are performed in real time~\footnote{As long as the resources allocated to each node are sufficient.}. It is therefore imperative that the constellation also be updated in real time. This argument is also made in $x$eoverse~\cite{kassem2024x}.

\begin{figure}
    \centering
    \includegraphics[width=0.45\textwidth]{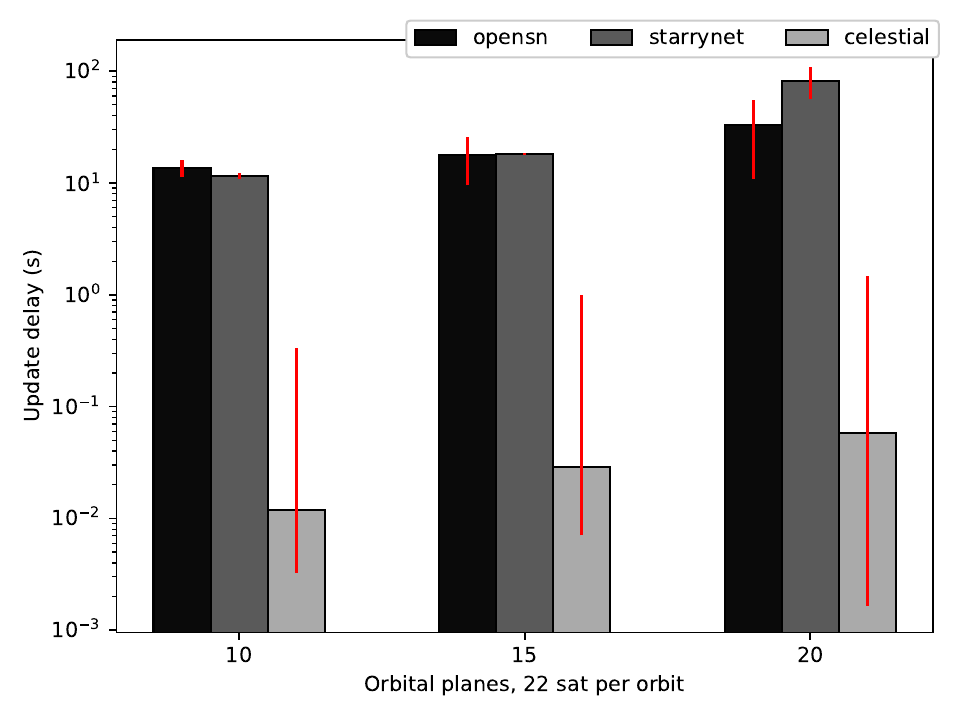}
    \caption{Delays incurred when updating the constellation characteristics, $y$ axis in $log$-scale.}
    \label{fig:update_delays}
\end{figure}

Constellation updates occur differently between the emulators. OpenSN uses an \textit{online} update procedure, where the positions of all satellites, the delays between them, and the links are updated at each emulation step. The delay calculations are computationally intensive and they introduce a significant overhead, compounded with the additional communication required by the disaggregated architecture. The result is that the simulation cannot maintain its step interval, which is 5s by default, and increases up to 55s at 20 orbital planes, as shown in Fig.~\ref{fig:update_delays}. StarryNet and Celestial adopt an \textit{offline-online} update procedure, where the calculation of delays and connectivity matrix are pre-computed (offline), and at simulation time (online) the network links are updated based on the pre-computed state. StarryNet performs both event-based updates, which mostly correspond to change of connectivity for ground stations, as well as periodic updates where the delays for the constellation are recomputed. Fig.~\ref{fig:update_delays} displays the latter, and the reason for the large values (almost 3x those of OpenSN for 20 orbital planes) is again that StarryNet (mainline version) employs CLI to interact with Docker and to configure the network links. Finally, the simplified network model of Celestial means that there are significantly fewer network connections that have to be updated. Celestial also uses \texttt{eBPF} to implement link traffic management, and that seems to be faster than the \texttt{tc} framework with \texttt{netem} employed by the other emulators.

\section{Observations and conclusions}
\label{sec:observations}

Emulation is the next best thing to working with real hardware. For it to be useful it needs to be realistic and also accessible. If measurements obtained through emulation are not representative of the real world they are useless. Also, if emulation requires as much resources as a real-world experiment, or if it takes too long to set up experiments, it loses its appeal. SplitNN~\cite{splitnn} makes the case that experiments that last several minutes should not take an hour to set up. Looking at the results presented in the evaluation of the three open-source satellite emulators shows that further improvements are required before the emulators can be truly useful.

From a fidelity point of view, all emulators show poor performance because they lack correct emulation of the lower-level communication. More effort is needed in the fidelity of link characteristics with more mature packet loss models and multi-user scheduling. Recent reverse-engineering studies of Starlink traffic management~\cite{tanveer2023making} and beam-forming~\cite{neinavaie2022unveiling} should be useful. In the current emulators the link model is embedded, perhaps a plugin-style architecture would facilitate improvements.

In terms of resource consumption, the node emulation approach obviously has a significant impact and it also affects bring-up time as well as constellation update time. The tested emulators show three options: Docker containers, Firecracker microVMs, and a custom container runtime. The best performing solution is that of StarryNet-dev, with the custom container runtime. StarryNet-dev does not currently use a root filesystem which makes it difficult to build complex applications for more mature scenarios, and ensure repeatability and consistency. It is however expected that with a root filesystem mounted (a "mount" namespace) the custom container runtime will be faster than a managed runtime such as Docker or Kubernetes, due to the lack of additional daemons.

The network emulation technology and method primarily affects the runtime overheads incurred when updating the constellation links. From the results, the Celestial \texttt{eBPF}-based solution seems to be the fastest, however it is difficult to gauge how effective it would be in managing a grid-like topology (as opposed to the star topology). It is also clear that the existing mechanisms for updating network interfaces at runtime are still insufficient for second-resolution constellation updates, when the size of the constellation increases. The constellation update times for all solutions grow with the network size, where they should actually be bounded. For node emulation, we can mitigate the scale with horizontal scaling over multiple machines. The same should be done for the constellation updates: the process should be parallelised and distributed over several machines.

The second issue related to networking is the computation of constellation state, which can be done before emulation (StarryNet and Celestial), or during (OpenSN). Considering the unstable overheads incurred by OpenSN at runtime would make a strong case for pre-computation. However, pre-computation results in very long experiment set-up times, even for short (several hours) simulations because the constellation state must be computed for each time step in advance. This can become prohibitive. Again, parallelisation and distribution are the solutions here, in which case the setup time can be bounded and then the constellation state update can be trace-driven, to reduce overheads.

Based on the results and the analysis of the satellite emulators, an ideal architecture is proposed for satellite emulators. The base technologies are:
\begin{itemize}
    \item custom or daemon-less container runtime (e.g. crun~\footnote{https://github.com/containers/crun}) with user-generated root filesystem
    \item network emulation based on \texttt{eBPF}
    \item parallelised pre-computation of constellation state
    \item parallelised constellation updates.
\end{itemize}
In terms of additional functionality, the following are recommended:
\begin{itemize}
    \item Visualisation interfaces are almost a requirement to be able to easily construct scenarios. A disconnected setup is better, where the scenario can be visualised separately to the emulation (as implemented by Celestial), rather than during the emulation (as in OpenSN).
    \item A bounding box feature can significantly reduce resource requirements and overheads for the majority of scenarios.
    \item The emulator should support heterogeneous satellite constellations, running distinct collections of applications on the nodes.
    \item The satellite constellation should be manually configurable in terms of number of orbital planes, arc and initial angle of ascending nodes; this can be valuable for conducting experiments with reduced resources as well as creating custom scenarios.
\end{itemize}

\balance
\bibliographystyle{IEEEtran}
\bibliography{sample-base}

\end{document}